\documentclass[12pt]{article}

\usepackage{amsfonts,amssymb,amsmath}

\usepackage{graphicx}
\DeclareGraphicsExtensions{.pdf,.eps,.jpg}

\begin{document}

\title{\bf Sensitivity of Photovoltaic Cells Efficiency to Initial Conditions in Various Aggregation Designs}
\author{Baharak Mohamad Jafari Navadel, Esfandyar Faizi, Baharam Ahansaz
\thanks{E-mail:bahramahansaz@gmail.com},
\\Jaber Jahanbin Sardroodi,
\\ {\small Physics Department, Azarbaijan Shahid Madani University, Tabriz, Iran}} \maketitle

\begin{abstract}
\noindent
It is thought that nature already exploits quantum mechanical properties to increase the efficiency of solar energy harvesting devices. So, the operation of these devices can be enhanced by clever design of a nanoscopic, quantum mechanical system where the quantum coherence plays a crucial role in this process.
In this investigation, we develop a donor-acceptor two level trap dipole model converging the key role of quantum coherence and aggregation effects along with different initial states. Our analysis reveals that quenching unwanted emissions is achievable by preparing the system in specific initial state under the effect of optimal spatial aggregation. Interestingly it is observed that characterizing aggregation-induced properties and quantum effects of bandgap engineering can increase the power enhancement up to $35.87\%$ compared with classical counterparts. This encouraging trend suggests a promising novel design aspect of nature-mimicking photovoltaic devices.

\end{abstract}

\newpage
\section{INTRODUCTION}
Photovoltaic cells (PV cells), also known as solar cells and photosynthesis, akin to classical heat engines, convert solar energy directly into electrical and chemical energy by the means of photovoltaic effect and a biological process, respectively. The pursuit of enhancing the energy conversion efficiency of PVs is the subject of extensive ongoing research. The performance of a solar energy harvesting device can be improved through the strategic design of a nanoscopic, quantum mechanical system. While thermodynamic principles establish the well-known Shockley-Queisser efficiency limit for classical photovoltaic devices, this limit can be surpassed by deliberately utilizing quantum interference to disrupt the detailed balance constraint $\cite{Tokihiro}$. It is hypothesized that nature leverages quantum mechanical properties to enhance the light-harvesting efficiency of photosynthesis. Prolonged quantum coherence has been observed in photosynthesis following laser excitation $\cite{Calhoun,Abramavicius,Panitchayangkoon,Harel,Hayes,Romero}$. This observation has attracted substantial interest in understanding how quantum coherence can be amplified in complex biological environments and its potential crucial role in efficient exciton transport processes $\cite{Mohseni,Plenio,Rebentrost,Zhu,Yeh}$. The FMO complex which connects the antenna to the reaction centre in the light harvesting apparatus of green sulfur bacteria, stands as the most extensively investigated system within this context $\cite{Fenna}$. Emulating photosynthesis offers a compelling avenue for enhancing the efficiency of contemporary solar cell technology $\cite{Blankenship}$. Establishing a connection between efficiency, functionality, and fortifying room-temperature quantum effects in these nanoscale systems could profoundly influence the design of future quantum-based nanotechnologies.

In a recent study, Dorfman and colleagues proposed an innovative solution to enhance light reactions in photocells $\cite{Dorfman}$. They analyzed these reactions as quantum heat engines (QHEs). By treating the light-to-charge conversion as a continuous Carnot-like cycle, they discovered that quantum coherence could significantly increase the photocurrent in a photocell based on photosynthetic reaction centers. Specifically, this boost amounted to at least $27\%$ when compared to an equivalent classical photocell. In their theoretical framework, the driving force behind this enhancement stems from the phenomenon of Fano interference $\cite{Scully1}$. Fano interference has been experimentally shown to allow optical systems to deviate from the thermodynamic detailed balance, which typically constrains the efficiency of light-harvesting devices. Notably, this violation of detailed balance was originally highlighted by Shockley and Queisser in their seminal work on the fundamental limits of semiconductor solar cells.
In this regard, Scully and his colleagues theoretically demonstrated that quantum coherence can enhance the performance of both a solar cell and a photosynthetic reaction center $\cite{Scully1,Scully2,Svidzinsky}$. Succeeding on the work of Scully et al., Creatore et al. proposed a biologically inspired photocell model enhanced by a delocalized dark quantum state involving two dipole-coupled donors $\cite{Creatore}$. Next, the authors of Ref. $\cite{Zhang}$ conducted an investigation on the scenario involving three coupled donors. Furthermore, Fruchtman et al. demonstrated that a photocell incorporating an asymmetric pair of coupled chromophores can surpass the performance of those containing a symmetric dimer or a pair of independent molecules $\cite{Fruchtman}$. The distinction between coherent and incoherent energy transfer has been extensively investigated in molecular crystals and aggregates. It is well-established that the interaction between exciton coupling and energetic disorder regulates the degree of exciton delocalization, which subsequently dictates the nature of transport. As excitons become more delocalized, coherent effects become increasingly significant.

In this paper we set out dipole coupled emitters and spatial nano-structure two-level trap required to fine-tune the emission properties such as power enhancement through preparing the system in different initial state and specific aggregation model. It is well established that emission properties are the reflection of interaction between optical molecular transition dipoles in collective excitations. We host dipole molecules of solar materials in H and J-aggregates producing delocalized states in order to supress recombination and increase photo-generated carriers. Theoretical  studies have shown that H-aggregate is more advantageous for charge transport due to the large $\pi-\pi$ overlap. Based on this scientific background, this research has provided systematic insights into the correlation between aggregation structure of dipole electron donors and emission induced properties causing. More interestingly, the system preparation in multiple initial states made a big improvement in power output enhancement intoducing unique, novel and more enhanced photocell model.

\section{MODEL DESCRIPTION}
In this paper, we develop the concept of a photosynthesis-inspired paradigm as QHE in order to increase the light-harvesting efficiency of quantum photocells.
The cyclic engine model considered here is a structure with three effective sites which mimics the photosynthetic reaction centers apparatus as in the primary model proposed by Dorfman $\cite{Dorfman}$. According to crystallography molecules in aggregates are most likely to be aligned collectively, not independently boosting individual optoelectronic properties in tunable light-matter interactions $\cite{Friend,Uoyama,Samuel,Tessler,Hu,Dong,Sirringhaus}$. The well-defined aggregation structure we consider here consists of two identical and initially uncoupled donor chromophores, ($D$, e.g., polymer-based material) which flank an acceptor molecule ($A$, e.g., fullerene-basedmaterial), as depicted in Figure 1. The optical excitation of donor configurations produced by solar radiation is modeled as two-level systems with the ground state $\vert b\rangle$ and the excited states $\vert a_{1}\rangle$ and $\vert a_{2}\rangle$.
The exciton dynamics of the donor aggregate structure is described by the Hamiltonian given by
\begin{eqnarray}\label{eq.0}
\mathcal{H}_{D}=\sum_{i=1,2}\hbar\omega_{i}\sigma_{i}^{+}\sigma_{i}^{-}+\sum_{i\neq j}J_{ij}(\sigma_{i}^{+}\sigma_{j}^{-}+\sigma_{i}^{-}\sigma_{j}^{+}),
\end{eqnarray}
where $\sigma_{i}^{+}=(\sigma_{i}^{-})^\dag=\vert a_{i}\rangle \langle b \vert$ with $i=1,2$ are the raising (lowering) operators.
By considering the donor molecules as a dipole moment with the optical transition dipole moment $\mathbf{\mu_{i}}=e \langle a_{i} \vert \mathbf{r} \vert b \rangle$ $(i=1,2)$, the electrostatic dipole-dipole coupling describes intermolecular interaction is given by
\begin{eqnarray}\label{eq.1}
  J_{12}=\dfrac{1}{4\pi\varepsilon\varepsilon_{0}}\bigg(\frac{\mathbf{\mu_{1}}.\mathbf{\mu_{2}}}{|\mathbf{R_{12}}|^{3}}-3\frac{(\mathbf{\mu_{1}}.\mathbf{R_{12}})(\mathbf{\mu_{2}}.
  \mathbf{R_{12}})}{|\mathbf{R_{12}}|^{5}}\bigg).
\end{eqnarray}
Consider dipole moment $\mathbf{\mu_{1}}$ located at position $\mathbf{R_{1}}$ and dipole moment $\mathbf{\mu_{2}}$ at position $\mathbf{R_{2}}$ coupled to electromagnetic radiation field through dipole transitions and the radius vector from $\mathbf{\mu_{1}}$ to $\mathbf{\mu_{2}}$ is $\mathbf{R_{12}}=\mathbf{R_{2}}-\mathbf{R_{1}}$. Typically, the strength of the interaction term $J_{12}$ is much weaker than the optical excitation energy and depends on the alignment of dipole moments. In a dimer system studied by Creatore et al. $\cite{Creatore}$, the donor dipole moment is always perpendicular to the radius vector $\mathbf{R_{12}}$, resulting in the vanishing of second term in Eq.(\ref{eq.1}). Thus, the electrostatic dipole-dipole coupling is given by $J_{12}\propto \mathbf{\mu_{1}}.\mathbf{\mu_{2}}=\mu_{1}\mu_{2}\mathrm{cos}\varphi$ where $\varphi$ is the angle between the two dipole moments. But with the assumption that the donor dipole moments are parallel ($\varphi=0$) and when the angle between them and the vertical axis denoted by $\theta$, the reduced form of Eq.(\ref{eq.1}) is given by
\begin{eqnarray}\label{eq.2}
J_{12}(\theta)=J_{12}^{0}\bigg(1-3\mathrm{cos}^{2}(\frac{\pi}{2}-\theta)\bigg).
\end{eqnarray}

Considering the schematic of reaction center as illustrated in Figure 1(a), initially both donor chromophores are quite optically active, synergistically facilitating the transition of excited electrons to the acceptor A. Dipole-dipole interaction occurs because of the electrostatic attraction between the positive and negative charges of the dipoles, impacting the electric field experienced by them, which in turn affects energy levels, transition probabilities, and other characteristics. The most important phenomena among these photophysical properties is the orbital overlap between adjacent donor molecules, resulting formation of new optically excitable states. Stable delocalized excited states, known as exciton states, are commonly observed in pigment-protein complexes $\cite{Van}$. These states play a crucial role in producing quantum interference effects that boost the photocurrent and the power of our QHE. In the presence of dipolar excitonic coupling $J_{12}$, the new donor eigenstates are formed as
\begin{eqnarray}\label{eq.3}
|x_{1}\rangle=\frac{1}{\sqrt{2}}(|a_{1}\rangle+|a_{2}\rangle),
\end{eqnarray}
which is symmetric/bright combinations of the uncoupled donor states and
\begin{eqnarray}\label{eq.4}
|x_{2}\rangle=\frac{1}{\sqrt{2}}(|a_{1}\rangle-|a_{2}\rangle),
\end{eqnarray}
which is antisymmetric/dark combinations of the uncoupled donor states. Also the corresponding energy eigenvalues are
\begin{eqnarray}\label{eq.5}
E_{x_{1}}=\frac{E_{1}+E_{2}}{2}+\sqrt{\frac{(E_{1}-E_{2})^{2}}{4}+J_{12}^{2}(\theta)},
\end{eqnarray}
and
\begin{eqnarray}\label{eq.6}
E_{x_{2}}=\frac{E_{1}+E_{2}}{2}-\sqrt{\frac{(E_{1}-E_{2})^{2}}{4}+J_{12}^{2}(\theta)},
\end{eqnarray}
where $E_{1}$ and $E_{2}$ represent the energies of the uncoupled donor states $\vert a_{1}\rangle$ and $\vert a_{2}\rangle$, respectively.
Due to the angle dependence of the mentioned energy eigenvalues, it is obvious that different aggregation structures will lead to distinct excited states and thus induce different luminescent properties. Herein, we assume two coupled donors ($D_{1}$ and $D_{2}$) are identical and degenerate, so the relationship between the angle and the Davidov energy splitting between the symmetric and antisymmetric states is given by $\bigtriangleup E(\theta)=2\vert J_{12}(\theta)\vert$.
In the following, we want to discuss about the aggregation structures named the H-aggregate and the J-aggregate structure.
In H-aggregate molecular structures, the two transition dipole moments are aligned in head-to-head manner with respect to vertical axis at $\theta=0$. Here, the symmetric state is higher than the antisymmetric state (see Figure 1(b)), so the optical transition is shifted to blue.
On the other hand, in J-aggregate molecular structures, the two transition dipole moments are aligned in head-to-tail manner with respect to vertical axis at $\theta=\pi /2$ and the antisymmetric state is higher than the symmetric state (see Figure 1(c)), so the optical transition is changed to red.
It also should be noted that the shift from the H-aggregate to the J-aggregate occurred at the approximately critical angle $\theta_{c}\thickapprox 35.26^{^{\circ}}$ with respect to vertical axis $\cite{Zhang}$. It is worth mentioning that classically the total dipole moment is always $2|\mu|$ when two dipole moments point to the same direction. But, the dipole moment of $|x_{1}\rangle$ is strengthened through constructive interference of the individual transition dipole matrix elements, $\mu_{x_{1}}=(\mu_{1}+\mu_{2})/\sqrt{2}=\sqrt{2}|\mu|$, whereas the dipole moment of $|x_{2}\rangle$ is annihilated due to destructive interference $\mu_{x_{2}}=0$. Therefore, the symmetric combination describes an optically active bright state with the optical transition rate $\gamma_{h}\propto |\mu_{x_{1}}|^{2}=2|\mu|^{2}$, which is doubled in comparison with an uncoupled donor case, while the antisymmetric combination describes an optically forbidden dark state.

Unlike previous studies in which the starting point of the cyclic light-emitting engine was considered in only ground state, in our new scheme, the initial state of our model is assumed to be in the following desired state given by
\begin{eqnarray}\label{eq.7}
  |\psi\rangle=\alpha|x_{1}\rangle+\beta|x_{2}\rangle.
\end{eqnarray}
Amazingly, we will find that under realistic constraints, different initial state effects surpass the performance of previously investigated systems and introduce an effective novel band gap engineering mechanism in order to harness quantum effects and coherent superposition advantages. Next, the excited electrons can be transferred to the acceptor molecule through electronic coupling and the emission of phonons, following the process described in $\cite{Dorfman}$. Subsequently, the excited electrons are utilized to perform work, resulting the charge-separated state $\vert \alpha \rangle$ decaying to the sub-stable state $\vert \beta \rangle$. Also, the recombination between the acceptor and the donor is also considered with a decay rate of $\chi \Gamma$, where $\chi$ is a dimensionless fraction. Ultimately, the state $\vert \beta \rangle$ undergoes decay, returning via $\Gamma_{c}$ to the charge neutral ground state $\vert b\rangle$ and complete the cycle.

We will now develop a kinetic scheme to describe the time evolution of the average occupations associated with this specific level structure.
As usual, we use the standard Born-Markov approximations: a weak interaction between an open quantum system and the environment and the extremely short correlation time of the environment, that is, no memory effect. So, we can obtain the master equation for the density matrix $\rho(t)$ of the photocell as follows
\begin{eqnarray}\label{eq.8}
\dfrac{\partial\rho}{\partial t}=(\dfrac{i}{\hbar}) [\rho,\mathcal{H}] +\mathcal{L}^{h}(\rho)+\mathcal{L}^{c}(\rho)+\mathcal{L}^{D}(\rho).
\end{eqnarray}
The Hamiltonian of the photocell is given by $\mathcal{H}=\mathcal{H}_{D}+\mathcal{H}_{A}$ where
$\mathcal{H}_{A}=\hbar\omega_{\alpha}\sigma_{\alpha}^{+}\sigma_{\alpha}^{-}+\hbar\omega_{\beta}\sigma_{\beta}^{+}\sigma_{\beta}^{-}$ is the Hamiltonian of the accepter and $\sigma_{\alpha}^{+}=(\sigma_{\alpha}^{-})^\dag=\vert \alpha\rangle \langle b \vert$ and $\sigma_{\beta}^{+}=(\sigma_{\beta}^{-})^\dag=\vert \beta\rangle \langle b \vert$ are the related raising (lowering) operators.
The Lindblad operator, denoted as $\mathcal{L}^{h}(\rho)$, encapsulates the interaction between the system and its surrounding hot bath, it is given by
\begin{eqnarray}\label{eq.9}
\mathcal{L}^{h}(\rho)=(\dfrac{\gamma_{h}}{2})(n_{h}+1)\bigg[2\zeta^{-}\rho\zeta^{+}-\zeta^{+}\zeta^{-}\rho-\rho\zeta^{+}\zeta^{-}\bigg]  +(\dfrac{\gamma_{h}}{2})n_{h}\bigg[2\zeta^{+}\rho\zeta^{-}-\zeta^{-}\zeta^{+}\rho-\rho\zeta^{-}\zeta^{+}\bigg],
\end{eqnarray}
where $\zeta^{+}=(\zeta^{-})^\dag=\vert x_{1}\rangle \langle b \vert$ is the corresponding raising (lowering) operator. In addition, $\gamma_{h}$ is the transition rate between ground state $|b\rangle$ and the bright state $|x_{1}\rangle$ and $n_{h}$ is the average thermal occupations at $T_{h}$ according to the Planck distribution $\cite{Carmichael,Breuer,Mandel}$.
Likewise, the Lindblad operator $\mathcal{L}^{c}(\rho)$, arising from the interaction with the cold bath can be written as follows
\begin{eqnarray}\label{eq.10}
\mathcal{L}^{c}(\rho)=\mathcal{L}^{c_{1}}(\rho)+\mathcal{L}^{c_{2}}(\rho)+\mathcal{L}^{c_{3}}(\rho),
\end{eqnarray}
where $\mathcal{L}^{c_{1}}(\rho)$ has the same form to Eq.(\ref{eq.9}) except replacing $\gamma_{h}\rightarrow \gamma_{x}$, $n_{h}\rightarrow n_{x}$ and
$\zeta^{+}\rightarrow \xi^{+}$ with $\xi^{+}=(\xi^{-})^\dag=\vert x_{1}\rangle \langle x_{2} \vert$. Similarly, the Lindblad operator $\mathcal{L}^{c_{2}}(\rho)$ ($\mathcal{L}^{c_{3}}(\rho)$) has the same form to Eq.(\ref{eq.9}) except replacing $\gamma_{h}\rightarrow \gamma_{c}$ ($\gamma_{h}\rightarrow \Gamma_{c}$), $n_{h}\rightarrow n_{c}$ ($n_{h}\rightarrow N_{c}$) and $\zeta^{+}\rightarrow \eta^{+}$ ($\zeta^{+}\rightarrow \tau^{+}$) where $\eta^{+}=(\eta^{-})^\dag=\vert x_{2}\rangle\langle\alpha \vert$ ($\tau^{+}=(\tau^{-})^\dag=\vert \beta \rangle \langle b \vert$) are the jump operators.
Correspondingly, $n_{x}$, $n_{c}$ and $N_{c}$ are the average thermal occupations at $T_{c}$ with energies $\Delta E=E_{x_{1}}-E_{x_{2}}$, $\Delta E=E_{x_{2}}-E_{\alpha}$ and $\Delta E=E_{\beta}-E_{b}$, respectively. The average thermal occupations at energies $\Delta E$ are given as follows
\begin{eqnarray}\label{eq.11}
n=\dfrac{1}{e^{\Delta E/k_{B}T_{c}}-1},
\end{eqnarray}
where $k_{B}$ stands for the Boltzmann constant. Ultimately, the Lindblad operator $\mathcal{L}^{D}(\rho)$ is expressed in a more detailed form as
\begin{eqnarray}\label{eq.12}
\mathcal{L}^{D}(\rho)=(\dfrac{\Gamma}{2})\bigg[2\lambda_{1}^{-}\rho\lambda_{1}^{+}-\lambda_{1}^{+}\lambda_{1}^{-}\rho-\rho\lambda_{1}^{+}\lambda_{1}^{-}\bigg] +(\dfrac{\chi\Gamma}{2})\bigg[2\lambda_{2}^{+}\rho \lambda_{2}^{-}-\lambda_{2}^{+}\lambda_{2}^{-}\rho-\rho \lambda_{2}^{+}\lambda_{2}^{-}\bigg].
\end{eqnarray}
In the mentioned equation, $\lambda_{1}^{+}=(\lambda_{1}^{-})^\dag=\vert \alpha\rangle \langle \beta \vert$ and $\lambda_{2}^{+}=(\lambda_{2}^{-})^\dag=\vert \alpha\rangle \langle b \vert$ are jump operators. The transition rate $\Gamma$ describes transitions from the state $\vert \alpha\rangle$ to the state $\vert \beta\rangle$. The parameter $\chi$ characterizes the recombination of charge carriers (electrons and holes) at the acceptor-donor interface effecting the current generation in a photocell. By assuming $\chi=0$ we analyze the scenario where no recombination occurs. This consideration aims to achieve the maximum power output from the photocell. The kinetics of light-matter interactions, following the Pauli master equation, guarantee completely positive populations. It is assumed that our proposed scheme functions as a quantum heat engine while being in thermal equilibrium with both hot and cold baths at the same time.

\section{RESULTS AND DISCUSSION}
Before the main discussion, we need to solve numerically the Pauli master equation Eq.(\ref{eq.8}) and obtain the transient and steady states.
In our calculation, we use the parameters listed in Table 1.
\begin{center}
\begin{tabular}{ |l|l| }
  \hline
  \multicolumn{2}{|c|}{Table 1. The used parameters.} \\
  \hline
  $E_{a_{1}} - E_{b}$ & 1.8 (eV) \\
  $E_{a_{2}} - E_{b}$ & 1.8 (eV) \\
  $E_{a_{1}} - E_{\alpha}$ & 0.2 (eV) \\
  $E_{a_{1}} - E_{\alpha}$ & 0.2 (eV) \\
  $E_{\beta} - E_{b}$ & 0.2 (eV) \\
  $\gamma_{h}=2\gamma_{1h}=2\gamma_{2h}$ & $1.24\times 10^{-6}$ \\
  $\gamma_{c}=2\gamma_{1c}=2\gamma_{2c}$ & $12\times 10^{-3}$ \\
  $\gamma_{x}$ & $25\times 10^{-3}$ \\
  $\Gamma_{c}$ & 0.0248 \\
  $J_{12}^{0}$ & 0.015 \\
  \hline
\end{tabular}
\end{center}

By incorporating the idea of photochemical current and voltage, we assign an effective current and voltage to the reaction center. The resistance of an external load is described by the electron decay rate $\Gamma$ from cathode to anode. The current flows from $\vert \alpha \rangle$ to $\vert \beta \rangle$ is expressed as $I =e \Gamma P_{\alpha}$, where $e$ represents the fundamental charge of an electron at steady state and the probability to find the system in state
$i(i=x_{1},x_{2},\alpha,\beta,b)$ is denoted as $P_{i}=\rho_{i,i}$, where $\rho_{i,i}$ is the diagonal elements of the density matrix.
It should be noted that the value of $\Gamma$ ranges from $\Gamma=0$ (the open circuit regime) to large $\Gamma$ (the short circuit regime).
The voltage across the solar cell is determined by the chemical potential variance between the mentioned two loads, $eV\equiv e(V_{\beta}-V_{\alpha})$=$\mu_{\alpha}-\mu_{\beta}$. By employing the Boltzmann distributions for levels $\vert \alpha \rangle$ and $\vert \beta \rangle$, where $P_{\alpha}=e^{-(E_{\alpha}-\mu_{\alpha})/K_{B}T_{c}}$ and $P_{\beta}=e^{-(E_{\beta}-\mu_{\beta})/K_{B}T_{c}}$ the solar cell's voltage is described in relation to energy levels and populations $\cite{Scully2,Creatore}$
\begin{eqnarray}\label{eq.13}
eV=E_{\alpha}-E_{\beta}+K_{B}T_{c}\ln\dfrac{P_{\alpha}}{P_{\beta}}.
\end{eqnarray}
When sunlight is absent, the system reaches thermal equilibrium with the phonon bath temperature, resulting in a vanishing voltage. Consequently, $V$ serves as a metric for quantifying the deviation from the thermal state characterized by temperature $T_{c}$. The rapid transfer rate of excited electrons to the acceptor results in a current proportional to the product of the electron charge and the generation rate, typically on the order of microamperes. It should be noted that the current and voltage are determined by analyzing the steady-state solutions of the master equation where we calculate the populations $\rho_{\alpha,\alpha}$
and $\rho_{\beta,\beta}$ at sufficiently long times.
Now, the power of the quantum photocell is given by
\begin{eqnarray}\label{eq.14}
P_{\mathrm{out}}=V I.
\end{eqnarray}
In this paper, in order to reach the desired power enhancement we integrate aggregation optimization, system preparation in different initial delocalized states, dark state protection and bandgap engineering. We discuss how aggregation becomes important and initializing the quantum photocell in a delocalized state gives rise to the big improvement of system performance compared with the previous studies. In this regard, we define the enhanced power output induced by the mentioned quantum effects as follows
\begin{eqnarray}\label{Equation4}
  PE=(\frac{P^{\mathrm{coupled}}_{\mathrm{out}}-P^{\mathrm{uocoupled}}_{\mathrm{out}}}{P^{\mathrm{uocoupled}}_{out}})\times100,
\end{eqnarray}
where $P^{\mathrm{coupled}}$ and $P^{\mathrm{uocoupled}}$ being the power output in the coupled ($J_{12}\neq0$) and uncoupled ($J_{12}=0$) case, respectively.

Figure 3 shows the power generated by the quantum photocell as a function of the coefficient $\beta$ in excitonically coupled $(J_{12}\neq 0)$ case for the H-aggregate configuration. Extracting power has been evaluated at room temperature $T_{c}=300$K. Figure 3(a) is plotted for the operation near the short circuit regime $\Gamma=0.1$eV and Figure 3(b) is plotted for the open circuit regime $\Gamma=0.001$eV. According to the Figure 3(a), when the initial state is prepared in the ground state $\vert b \rangle$, the power output always is higher compared to the general delocalized initial state $\vert \psi \rangle$. Meanwhile, this result is completely reversed in the case of $\Gamma=0.001$eV. The maximum power output $P_{\mathrm{out}}^{\mathrm{max}}=4334.3 \mu W$ of the studied photocell is obtained for $\beta=0$ representing the initial state $\vert x_{1} \rangle$ in the H-aggregation. Also, the production of output power values in descending order according to the preparation of the specific initial states in H-aggregation is: $P_{x_{1}}>P_{a_{1}}=P_{a_{2}}>P_{x_{2}}$. In contrast to the previous claim, considering the H-aggregation type for $\Gamma=0.001 eV$, when the system is initialized in the general delocalized state $\vert \psi \rangle$ the resulting power output is greater than that obtained from the ground state $\vert b \rangle$. In contrast to the previous case, the power output of the photocell reaches its minimum value for the initial state $\vert x_{1} \rangle$ and the maximum power corresponds to the initial state $\vert x_{2} \rangle$.

Figure 4 presents the generated power as a function of the coefficient $\beta$ in excitonically coupled $(J_{12}\neq 0)$ case for the J-aggregate configuration.
As depicted in Figure 4(a,b), the analysis examines two distinct values near the short circuit regime ($\Gamma=0.1$eV) and near the open circuit regime ($\Gamma=0.001$eV), respectively. According to Figure 4(a), the power output is greater when the initial state is prepared in the ground state $\vert b\rangle$ compared to the prepared initial state in the general delocalized state $\vert \psi \rangle$. Conversely, this outcome is entirely reversed in the case with $\Gamma=0.001$ eV. The minimum power output obtained using the discussed quantum effects across various configurations and by varying the $\Gamma$ parameter is $P_{\mathrm{out}}^{\mathrm{min}}=78.34 \mu W$, occurring with the initial state $\vert x_{2}\rangle$ in the J-aggregation and for $\Gamma=0.001$eV. As observed in Figure 4(a), the descending order of output power values based on the specific initial states preparation in the J-aggregate configuration is as follows: $P_{x_{2}}>P_{a_{1}}=P_{a_{2}}>P_{x_{1}}$. Similar to Figure 3(b) when the system is initialized in the general delocalized state $\vert \psi \rangle$ the resulting power output is always greater than that obtained from the ground state $\vert b \rangle$. In this scenario, the power output of the photocell reaches its minimum for the initial state $\vert x_{2}\rangle$. Conversely, the maximum power output corresponds to the initial state $\vert x_{1}\rangle$ at $\beta=0$, demonstrating the peak performance.

The angle dependence of the power enhancement $\mathrm{PE}$ near the short and open circuit regimes is illustrated in Figure 5.
As shown in Figure 5(a), it is obvious that the power enhancement is largely independent of the system's initial state, as the graphs are nearly identical and overlaped. In fact, the power enhancement does not significantly depend on the initial state and starting point of the system.
Moreover, the enhancement of power is affected by the aggregation of dipole moments, with the optimal condition occurring at the H-aggregate or when $\theta=0$. As $\theta$ increases, the power enhancement decreases, hitting its minimum value in the J-aggregate state. At the critical angle $\theta_{c}$ (transition from the H-aggregate phase to the J-aggregate), $J_{12}$ becomes zero, indicating the possibility of an uncoupled regime. Therefore, the uncoupled power is equal to the power of the coupled regime, resulting in zero power enhancement. In the following, we investigate the angle dependency of the $\mathrm{PE}$ with the used parameter $\Gamma=0.001$eV, as depicted in Figure 5(b). In this scenario, the $\mathrm{PE}$ interestingly exhibits a significant reliance on the initial state of the system. For the H-aggregate case, the $\mathrm{PE}$ demonstrates positive values and it follows a ascending trend for the different initial $\beta=(0, 2^{-1/2},1)$ with approximately values of $17.64\%$, $26.75\%$ and $35.87\%$, respectively. It drops zero at magic angle $\theta=\theta_{c}$ and after crossing the critical angle, the $\mathrm{PE}$ takes negative values and ultimately it approximately reaches $-42.19\%$, $-64.64\%$ and $-87.08\%$ in the J aggregate configuration of dipoles. Overall, we conclude that when $\theta < \theta_{c}$, the power of coupled case in higher than the power of uncoupled case and when $\theta > \theta_{c}$, the power of coupled case in lower than the power of uncoupled case.

To give a more clear physical reason for our results in Figure 5, we emphasize that for the H-aggregate, the optimal state occurs at $\theta=0$, resulting in maximum power enhancement due to constructive dipole coupling. As $\theta$ increases from this optimal alignment, the power enhancement decreases, reflecting a transition from constructive to destructive interference of dipole interactions. Conversely, in the J-aggregates, the power enhancement reaches its lowest value. This is due to the increased destructive interference as the dipoles misaligned. Moreover, it is worth noting that for the H-aggregate case ($\theta < \theta_{c}$), the bright state $|x_{1}\rangle$ lies at a higher energy than the dark state $|x_{2}\rangle$ and for the J-aggregate case ($\theta > \theta_{c}$), the bright state $|x_{1}\rangle$ lies at a lower energy than the dark state $|x_{2}\rangle$. Also, by comparing the energy splitting between the bright and dark states for the H and J-aggregates, we have $\Delta E^{\mathrm{J-aggregate}}=2\Delta E^{\mathrm{H-aggregate}}$. Consequently, the transition from the donor to the acceptor for the J-aggregate case is very low and the current enhancement is negative because an electron in the bright state jumps to the dark state through only the stimulated absorption $\gamma_{x}n_x$ of thermal phonons for the J-aggregate case while this transition occurs through the stimulated and spontaneous emissions of thermal phonons $\gamma_{x}(1 + n_x)$ for the H-aggregate case.

\section{CONCLUSIONS}
In summary, the importance of the initial-state dependence for a photovoltaic cell efficiency under different aggregation and realistic constraints has been investigated near the short and open circuit regimes. We have considered dipole molecules of solar materials producing delocalized states and amazingly found that the power output with the initially prepared general delocalized excited state can surpass the performance of previously studied systems with the initially prepared ground state near the open circuit regime and in the both H and J aggregation. In addition, our findings indicated that our considered photocell can result in a notable enhancement of power output by about $35.87\%$ in a coherent coupled dipole system compared with the uncoupled dipoles in the H aggregate and near the open circuit regime. This study provides valuable insights into the improvement in PV cells conversion efficiency offering a foundation for future investigations and practical applications in light-harvesting industry.

\section*{Data availability}
The data of the present study are available from the corresponding author upon a reasonable request.

\section*{Conflict of interest}
The authors declare that they have no conflict of interest.

\newpage
\begin{figure}
        \centering{
        \qquad $\mathbf{(a)}$ \qquad\qquad \qquad\quad\qquad\qquad\qquad $\mathbf{(b)}$\\{
        \includegraphics[width=3.2in]{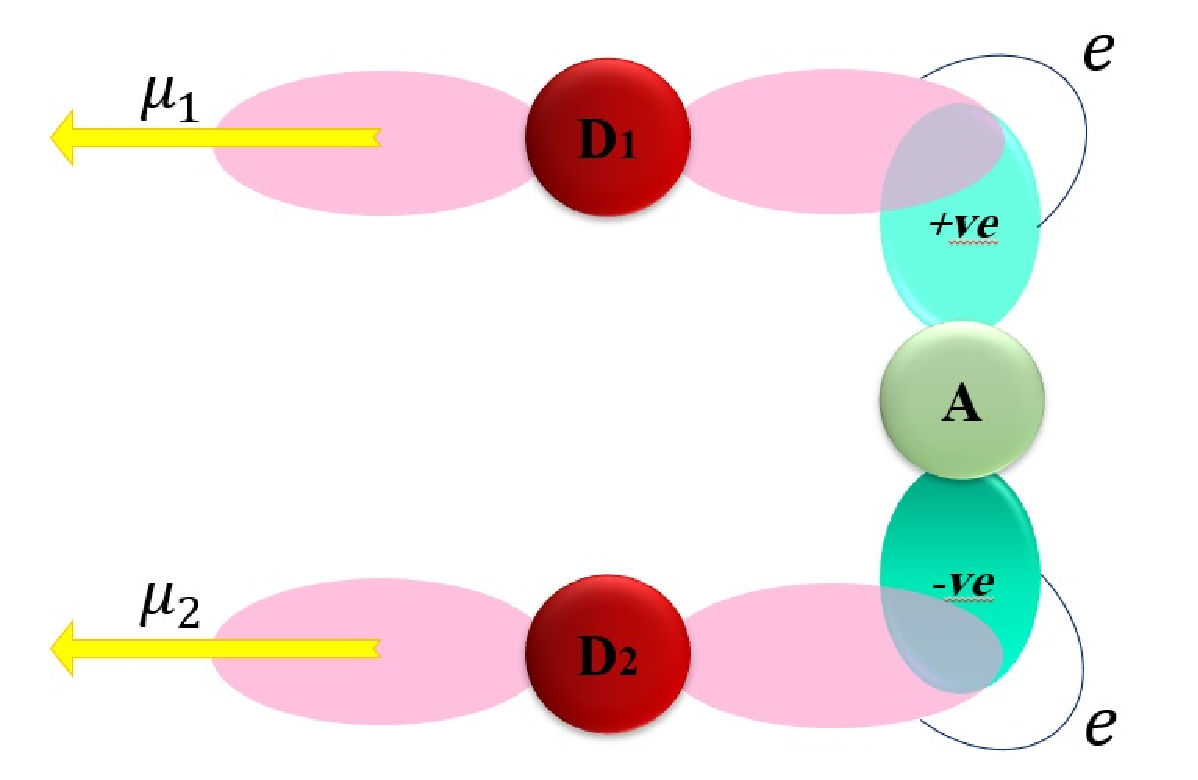}
        \label{fig:first_sub}
    }{
        \includegraphics[width=2.2in]{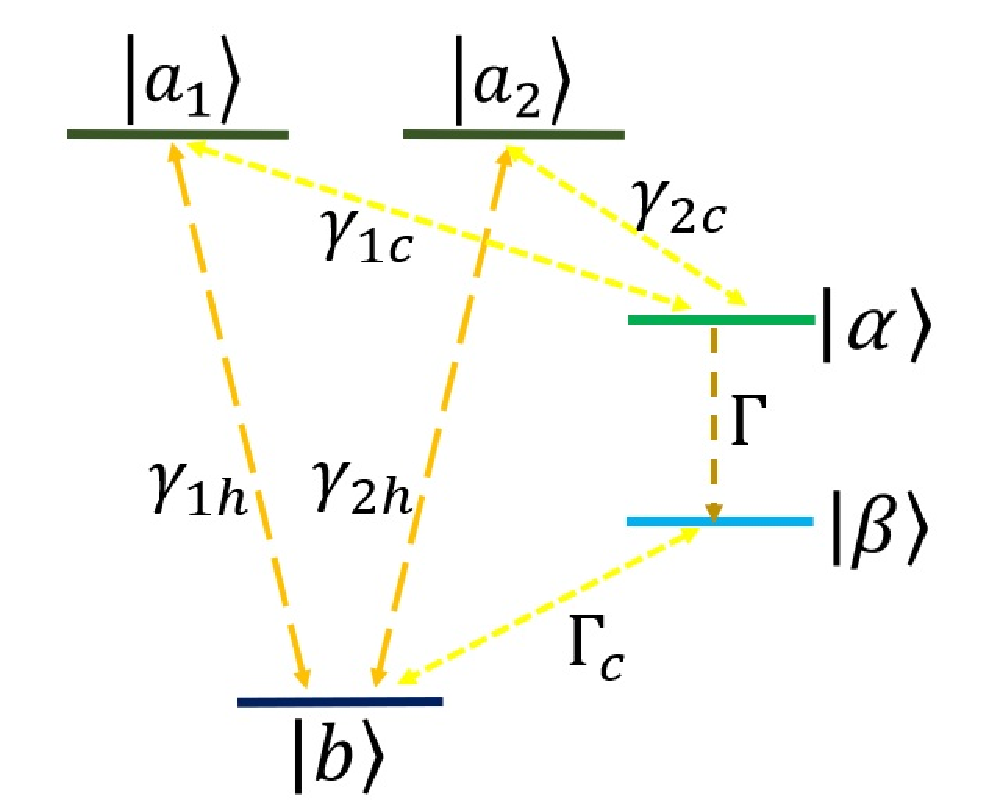}
        \label{fig:second_sub}
    }\\ \par \qquad $\mathbf{(c)}$ \qquad\qquad \qquad\quad\qquad\qquad\qquad $\mathbf{(d)}$\\{
        \includegraphics[width=3.2in]{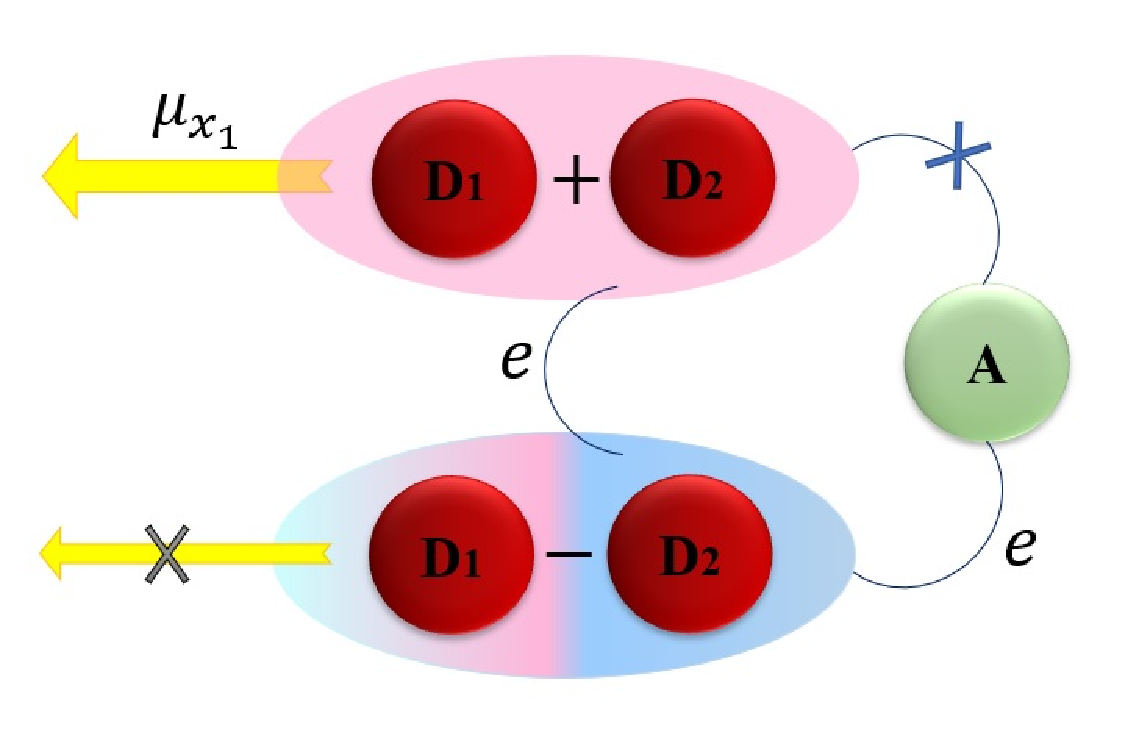}
        \label{fig:first_sub}
    }{
        \includegraphics[width=2.2in]{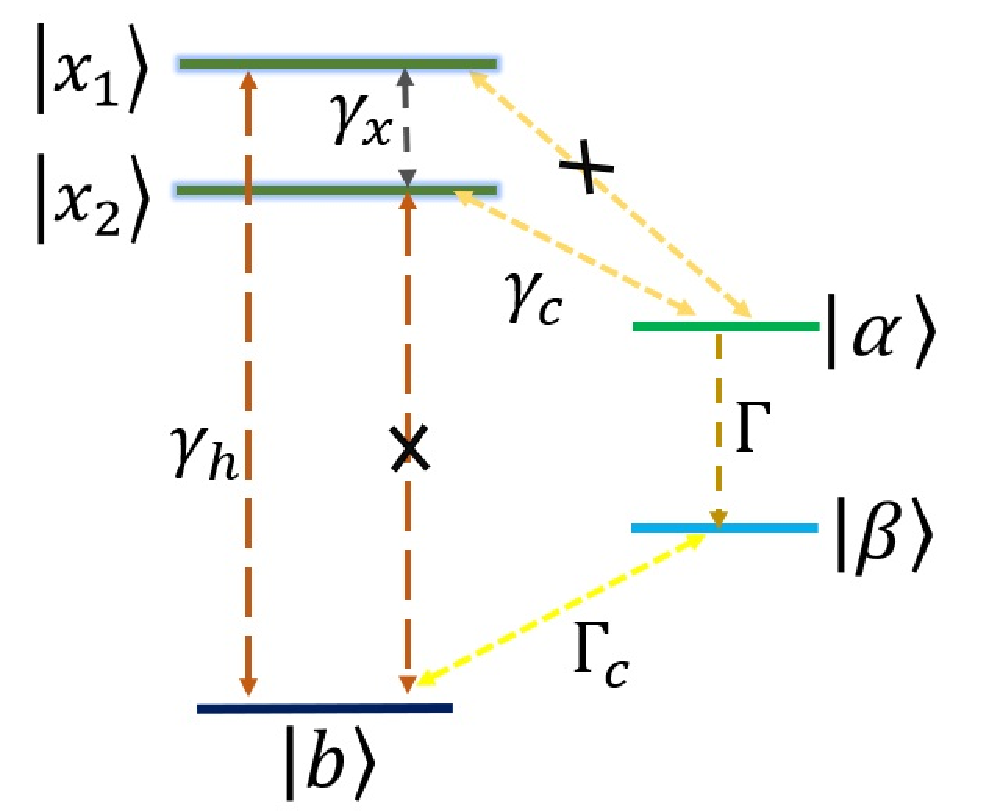}
        \label{fig:second_sub}
    }\\ \par \qquad $\mathbf{(e)}$ \qquad\qquad \qquad\quad\qquad\qquad\qquad $\mathbf{(f)}$\\{
        \includegraphics[width=3.2in]{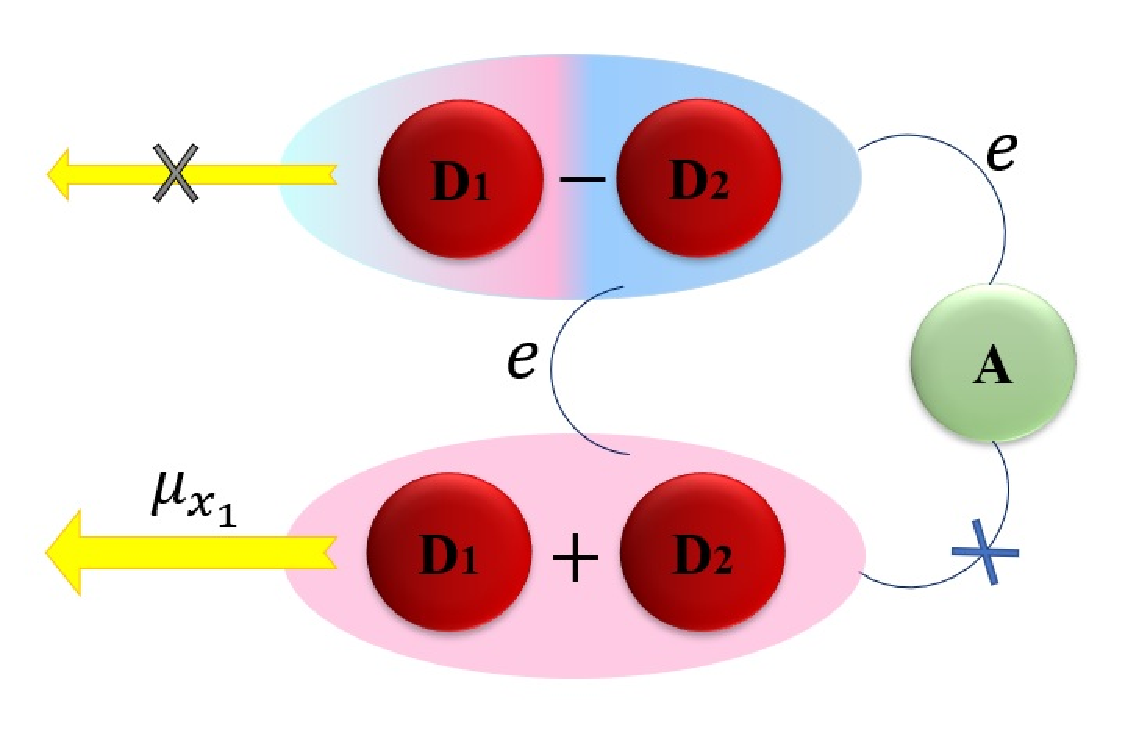}
        \label{fig:first_sub}
    }{
        \includegraphics[width=2.2in]{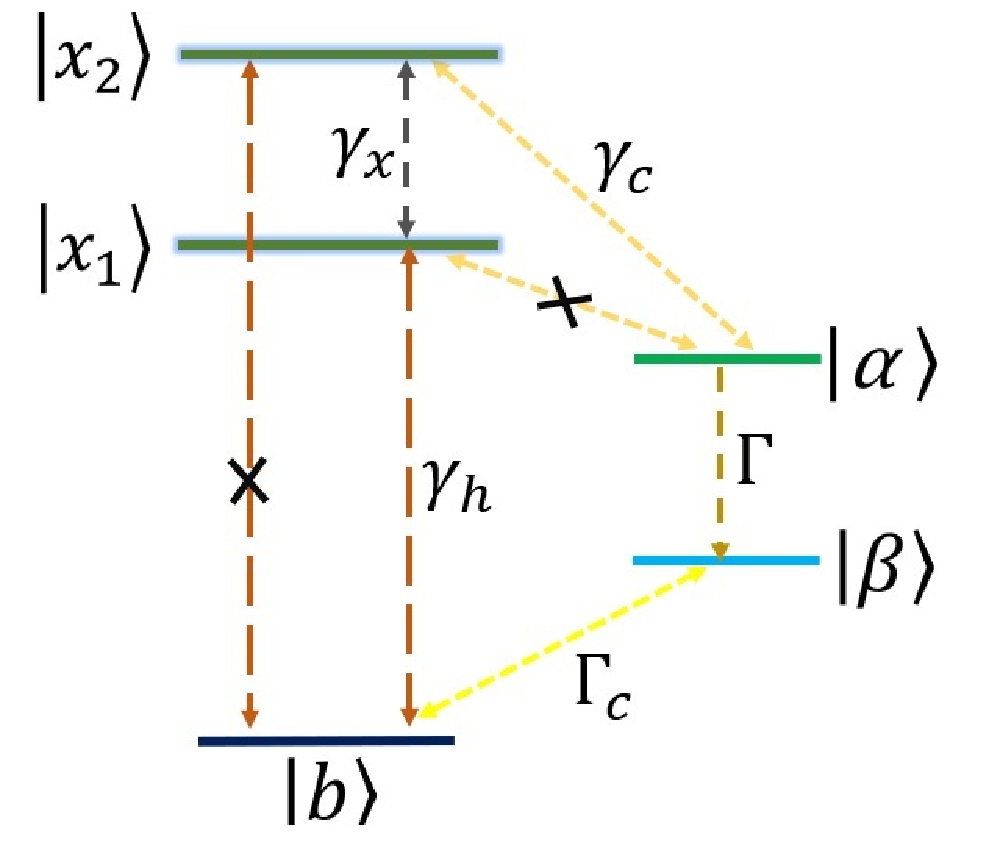}
        \label{fig:second_sub}
    }
        }
    \caption{Schematic of our PV cell with the energy level diagrams. In (a), uncoupled case, both donors $D_{1}$ and $D_{2}$ exhibit optical activity and synergistically facilitate the transfer of excited electrons to the acceptor A. The red and blue shaded/pale regions encircling the molecules illustrate the molecular orbitals, which depict the spatial electron density distribution. In (c) and (e), coupled cases, the interaction between $D_{1}$ and $D_{2}$ leads to the formation of a coupled system which generates new eigenstates superposed of the uncoupled donor states $\vert a_{1}\rangle$ and $\vert a_{2}\rangle$. (b), (d) and (f) show the level structure and electron paths through bright and dark states of uncoupled, H-aggregate and J-aggregate, respectively.}
    \end{figure}

\newpage
\begin{figure}
        \centering{
        \includegraphics[width=4in]{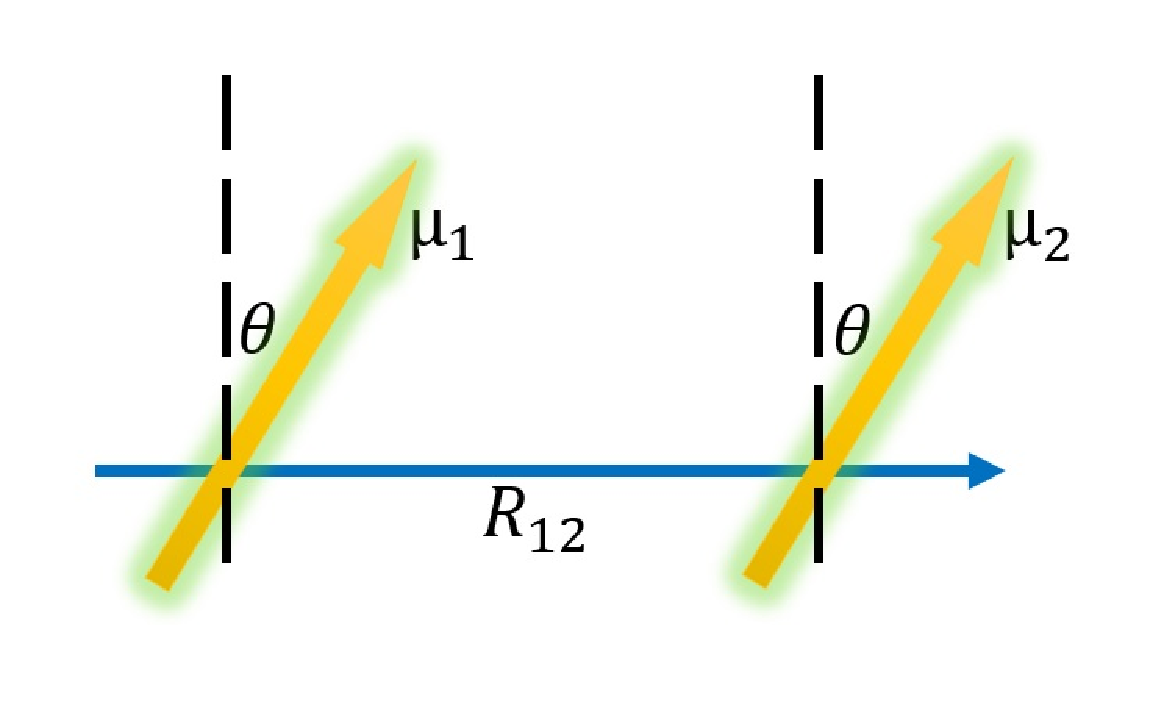}
        \label{fig:first_sub}
        }
    \caption{The alignment of two parallel dipole moments $\mu_{1}$ and $\mu_{2}$ is depicted as yellow arrows. Here, the angle $\theta$ is measured with respect to the vertical axis.}
    \end{figure}

\newpage
\begin{figure}
        \centering (a)\\{
        \includegraphics[width=4.4in]{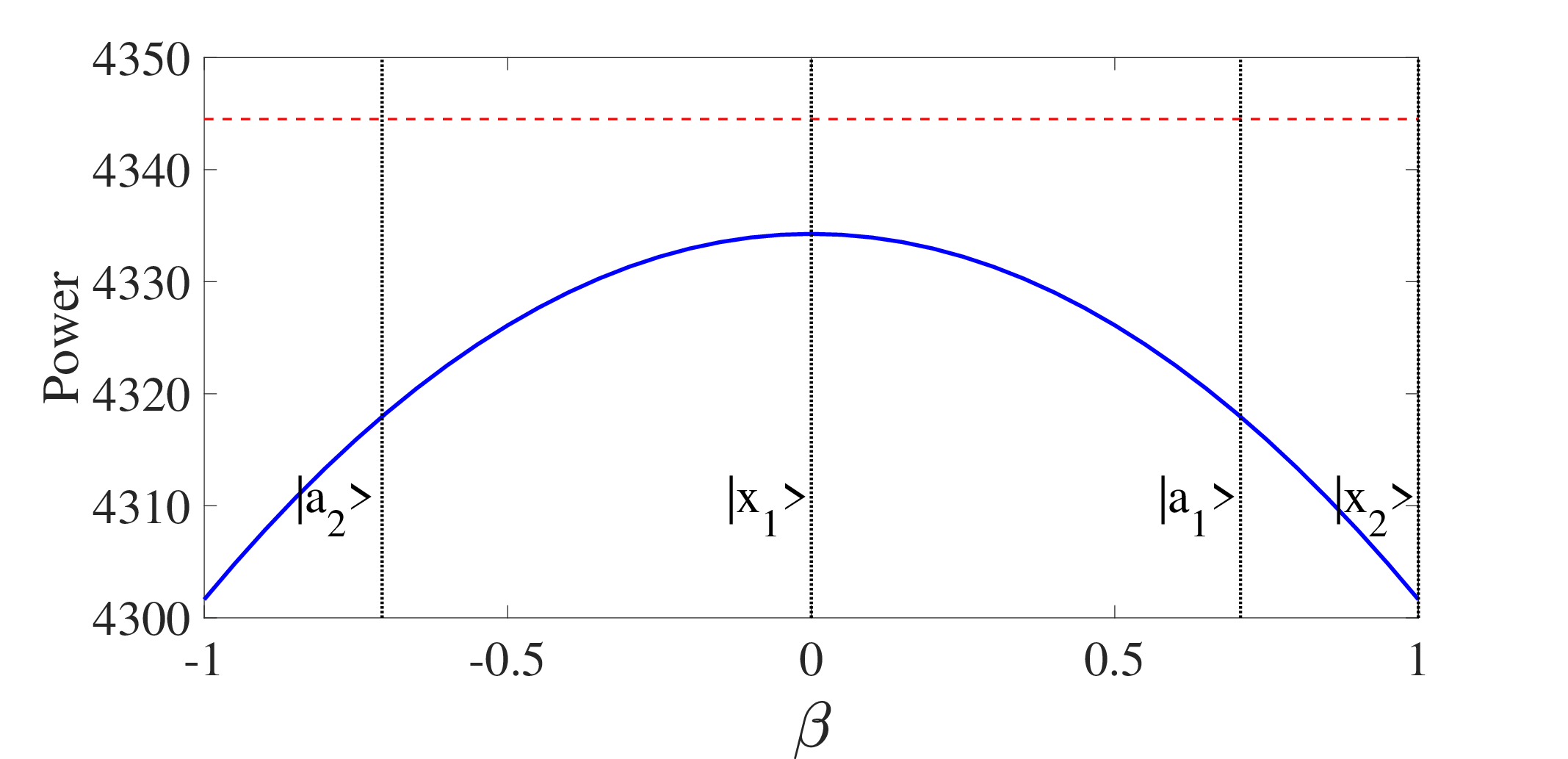}
        \label{fig:first_sub}
    }\\ \centering (b)\\{
        \includegraphics[width=4.4in]{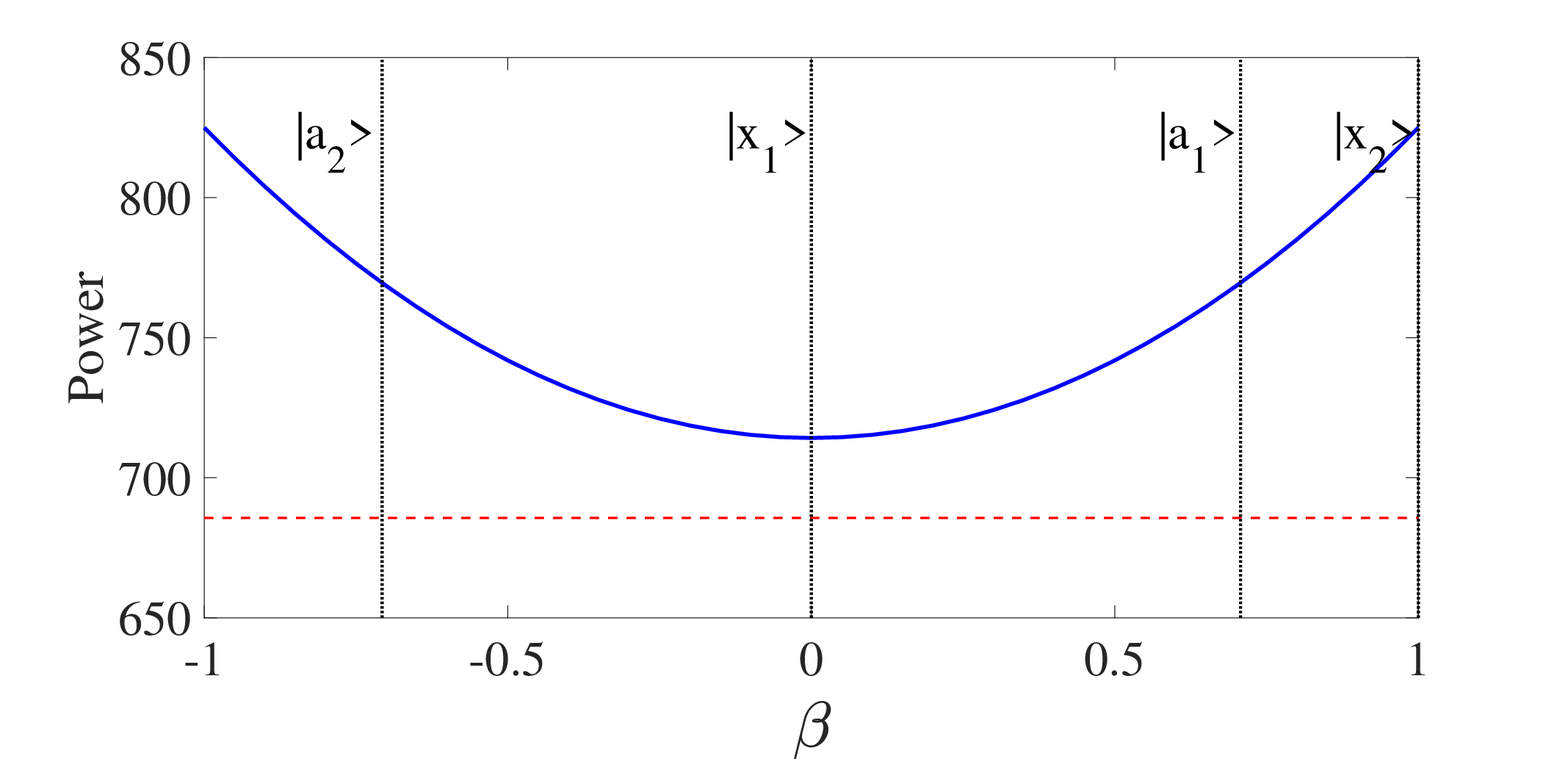}
        \label{fig:first_sub}
    }
    \caption{Power generated in PV cell as a function of the coefficient $\beta$ at room temprature when system reaches steady state operation in the presence of coherent dipole coupling ($J_{12}\neq 0$) with the H-aggregate configuration for (a) $\Gamma=0.1$eV and (b) $\Gamma=0.001$eV. The red dashed line is a benchmark line denoted for comparison of the generated power in all delocalized states with the initially prepared ground state $\vert b \rangle$.}
    \end{figure}

\newpage
\begin{figure}
        \centering (a)\\{
        \includegraphics[width=4.4in]{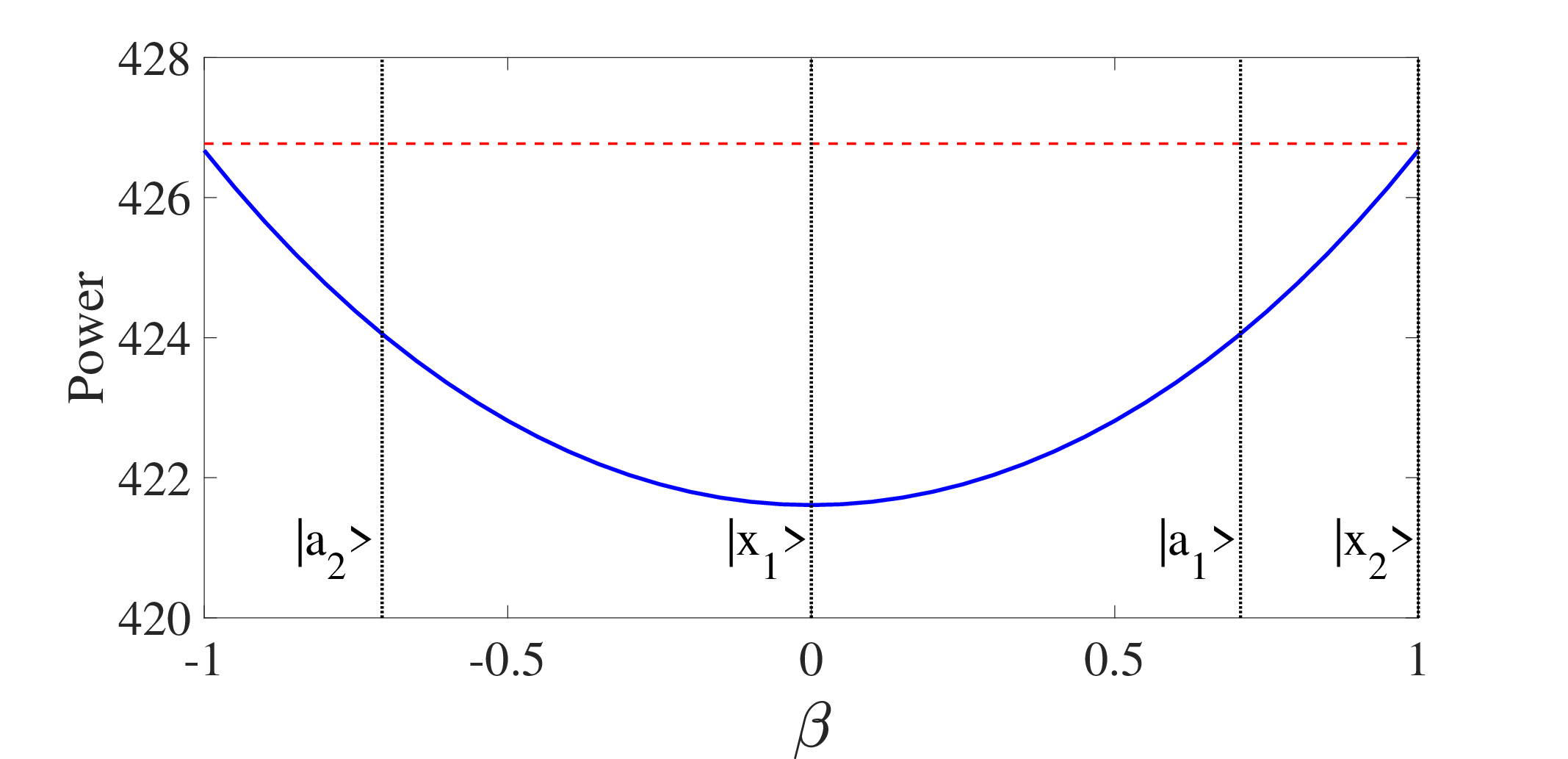}
        \label{fig:first_sub}
    }\\ \centering (b)\\{
        \includegraphics[width=4.4in]{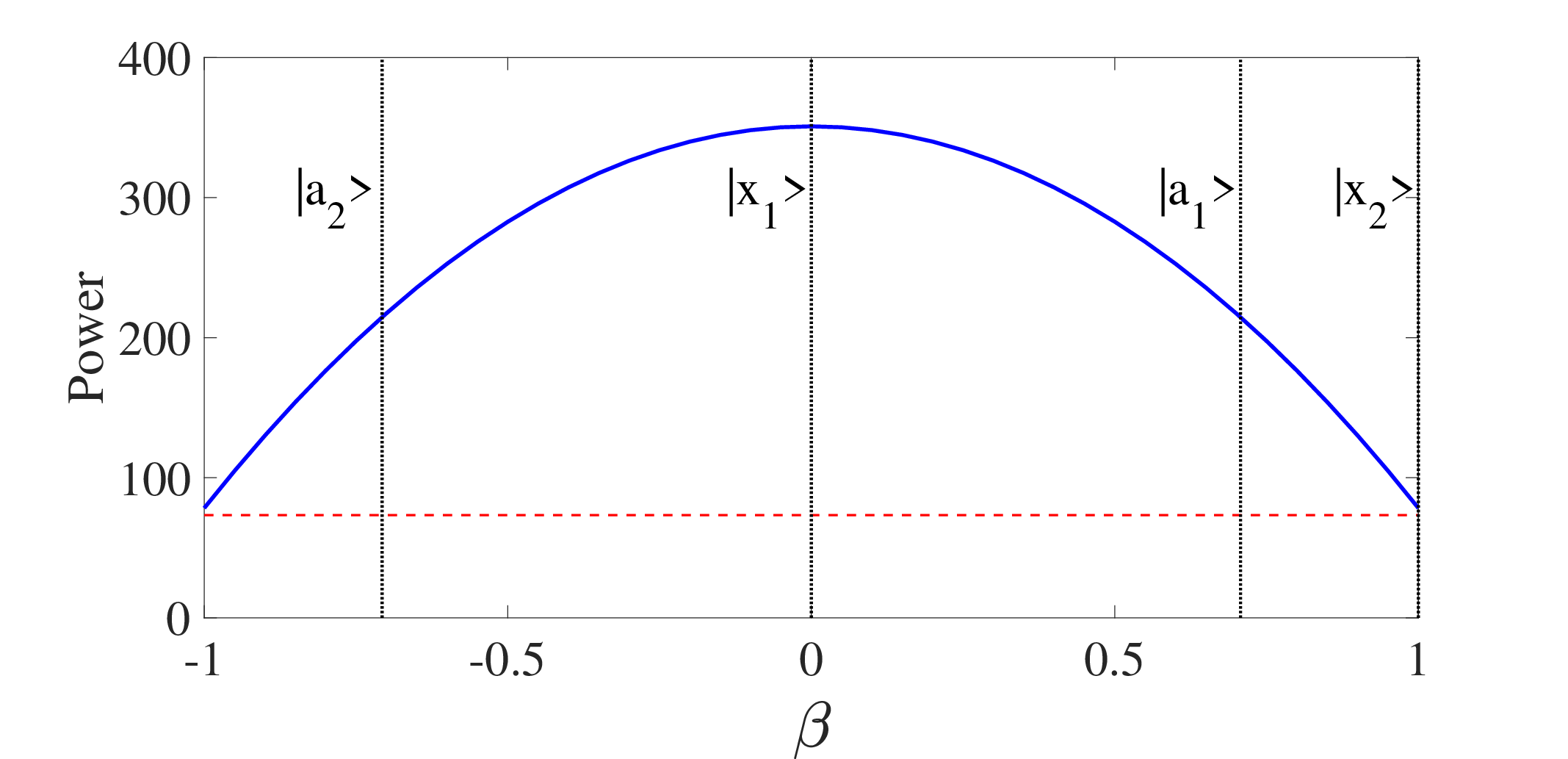}
        \label{fig:first_sub}
    }
    \caption{Power generated in PV cell as a function of the coefficient $\beta$ at room temprature when system reaches steady state operation in the presence of coherent dipole coupling ($J_{12}\neq 0$) with the J-aggregate configuration for (a) $\Gamma=0.1$eV and (b) $\Gamma=0.001$eV. The red dashed line is a benchmark line denoted for comparison of the generated power in all delocalized states with the initially prepared ground state $\vert b \rangle$.}
    \end{figure}

\newpage
\begin{figure}
        \centering (a) \qquad \quad\qquad\qquad\qquad\qquad\qquad\qquad (b)\\
        \centering{
        \includegraphics[width=3in]{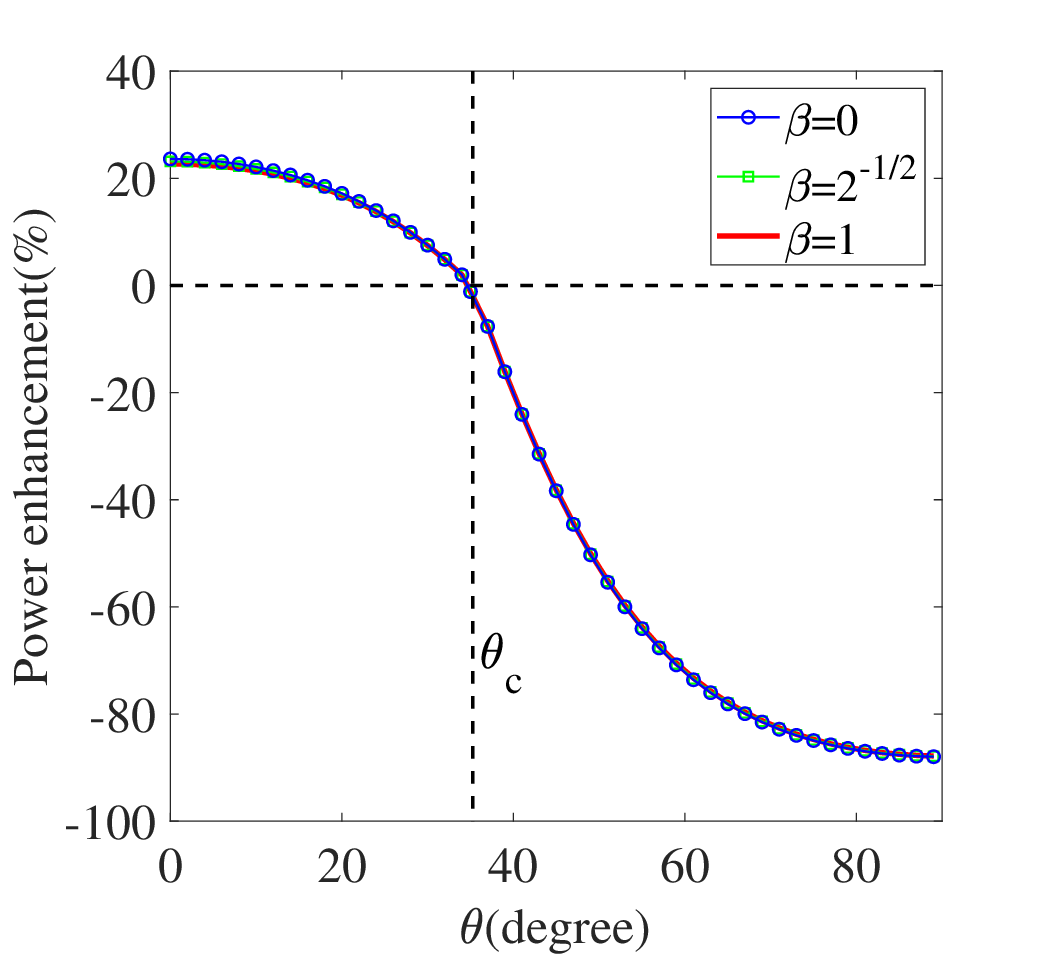}
        \label{fig:first_sub}
    }\centering{
        \includegraphics[width=3in]{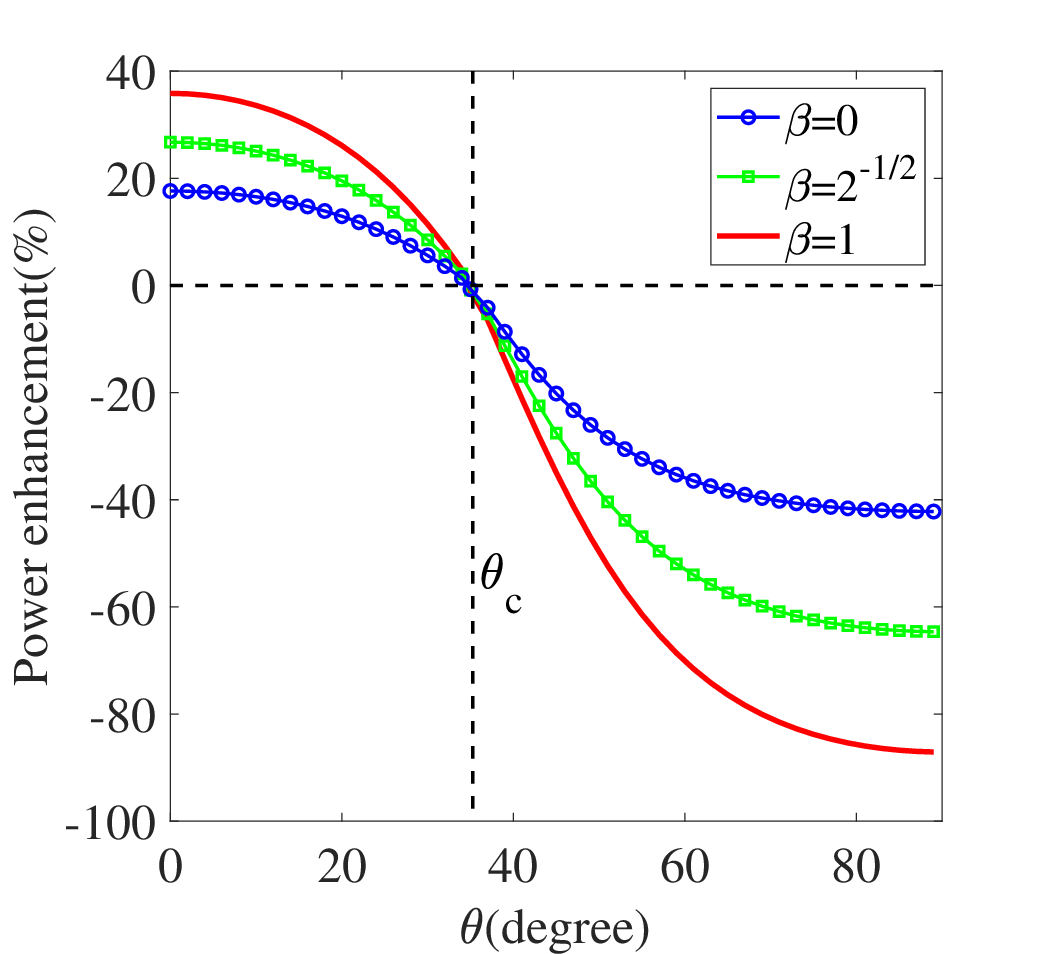}
        \label{fig:second_sub}
    }
    \caption{Power enhancement as a function of the angle $\theta$ with respect to the vertical axis for (a) $\Gamma=0.1$eV and (b) $\Gamma=0.001$eV. Blue circle line denotes power proportional to $\beta=0$, green square line denotes $\beta=2^{-1/2}$ and red line denotes $\beta=1$ representing the set up prepared in initial states $\vert x_{1} \rangle$, $\vert a_{1}\rangle$ and $\vert x_{2} \rangle$, respectively. The black arrow indicates the specific angle known as the magic angle $\theta_{c}\approx 35.26^{\circ}$.}
    \end{figure}

\end{document}